# A Novel Method for Stock Forecasting based on Fuzzy Time Series Combined with the Longest Common/Repeated Sub-sequence

He-Wen Chen, Zih-Ci Wang, Shu-Yu Kuo, Yao-Hsin Chou
Dept. of Computer Science and Information Engineering
National Chi-Nan University
No.1 University Rd., Puli 54561, Taiwan
Email : yhchou@ncnu.edu.tw

*Abstract*—Stock price forecasting is an important issue for investors since extreme accuracy in forecasting can bring about high profits. Fuzzy Time Series (FTS) and Longest Common/Repeated Sub-sequence (LCS/LRS) are two important issues for forecasting prices. However, to the best of our knowledge, there are no significant studies using LCS/LRS to predict stock prices. It is impossible that prices stay exactly the same as historic prices. Therefore, this paper proposes a state-of-the-art method which combines FTS and LCS/LRS to predict stock prices. This method is based on the principle that history will repeat itself. It uses different interval lengths in FTS to fuzzify the prices, and LCS/LRS to look for the same pattern in the historical prices to predict future stock prices. In the experiment, we examine various intervals of fuzzy time sets in order to achieve high prediction accuracy. The proposed method outperforms traditional methods in terms of prediction accuracy and, furthermore, it is easy to implement.

## I. INTRODUCTION

The stock market plays an important role in any country's economic system because it houses investment targets which are easy to buy/sell and have complete information that is available to the public. The properties mentioned above cause numerous people to invest in the stock market in the hopes of making a profit. In this world, there are many stock markets, but only Taiwan's stock market is special; it is not only affected by economically powerful countries, such as America, China, Japan, European countries, etc., but also features a great system where stock prices cannot be higher/less than 7% when compared with the previous day's prices. Moreover, people from all around the world are able to invest in the Taiwanese stock market.

Several research studies have focused on issues that are pertinent to the stock market, such as price forecasting, deciding on trading points and stock selection. Price forecasting is typically the foundation of all of these studies. If the results of forecasting systems are made accurate, then decisions regarding trading points and stock selection will result in the acquisition of greater income. Thus numerous mathematical tools have been applied, such as Evolution computation (EC) [1], linear and multi-linear regression (LR, MLR) [2], Artificial neural networks (ANNs) [3], FTS [4], etc.

In addition, some researchers [5] recommend a method which uses LCS/LRS to predict stock prices. This method is usually applied to a DNA sequence in order to recognize the particular gene fragment. Most researchers believe that history repeats itself. And so, as a hypothesis, LCS/LRS is a powerful method that is able to identify repeated fragments in the past, and provide more information for forecasting in the future.

However, it is difficult to match historic prices accurately, particularly in terms of the second decimal place, because prices are always changing over time. Fortunately, a mathematical tool called "FTS" can be used to solve the mapping issue as it can blur the price and so results in higher success rates in terms of mapping.

However, FTS has its own disadvantages. One relates to interval design and the other relates to one/higher order fuzzy logical relationships. Previous research has used the highest and lowest prices of historic data to serve as the domain range of FTS, but if the future price is higher or lower than the domain range, then this causes prediction inaccuracy. We therefore use the percentage of rises and falls in price as a domain range, this problem can be avoided. One/higher order fuzzy logical relationships also cannot represent the actual patterns shown in history. This research therefore proposes a novel concept, which combines the features of "LCS/LRS" and "FTS", and forecasts Taiwan's stock exchange capitalization weighted stock index (TAIEX) and Taiwan's Top 50 Exchange Traded Funds (ETF) in order to approximately represent trends in Taiwan's stock market. This combination allows us to discard the drawbacks of FTS. The experimental results show that the proposed method is more effective than the FTS, and, further, demonstrates the drawbacks of FTS.

This article defines LCS/LRS and introduces some related works in Sec. 0 and 0, respectively. Sec. IV discusses the proposed method. Subsequently, some results of FTS and the proposed method are analyzed in Sec. 0, and we provide a conclusion in Sec. VI.

## II. BACKGROUND

LCS/LRS is a string tool. The original LCS is given certain sequences and identifies the common aspects in all of these sequences. For instance, from two sequences $S_1$=abcde and $S_2$=bdscjkde, the aim is to identify the common part which underlies $S_2$, "de"; an algorithm is usually used to compare the common part in multiple gene sequences. And the target of the original LRS is to locate the repeated subsequence of a sequence; for example, S is abcklesabcgkels and the repeated sequence is abc.

In this research, the definition of the repeated subsequence is similar to the original LRS. The principle is that by giving another sequence and order numbering, we are able to identify the repeated subsequences as well as the next letter in the sequence, for example, S1=abcklesabcgkels, S2=jeusabc and the order is 3; the repeated subsequences are "abc,k" and "abc,g" respectively. The number of repeated subsequences is influenced by the number of the order, for instance, if the number of the order is 4, from example above, there will be no subsequence similar to "sabc". The mapping is as follows:

Given a sequence to represent the historical data, which is $S_1$="$A_iA_i…A_i$", where i ∈ {1, 2, … , k}, and $k$ is the number interval. And another sequence $S_2$="…$A_i^3 A_i^2 A_i^1$", the algorithm sets a counter to 1 at first, and mapping starts from the last state, which is $A_i^1$ where the superscript is the same as the counter. It then identifies all of the states that are the same as $A_i^1$ where the $i$ values are the same. The counter is then added by 1, and the returned LCS/LRS must correspond to $A_i^2 A_i^1$, and so on. The algorithm stops when the counter is equal to the number of the order or the returned LCS/LRS is NULL.

For example, the number of the interval is 3, and the history data is $S_1$="$A_1A_3A_2A_2A_1A_3A_1$". Another sequence is $S_2$="$A_2A_1A_3$". The counter=1, and the number of the order is 3. The algorithm starts from the last state at $S_1$ which is "$A_3$" and returns two LCS/LRS which are "$A_3A_2$" and "$A_3A_1$" in the first round, respectively. After this, the counter is added by 1 and also returns two LCS/LRS, which are "$A_1A_3A_2$" and "$A_1A_3A_1$" in the second round, respectively. In the third round, the counter is equal to 3 and only one LCS/LRS is returned, which is "$A_2A_1A_3A_1$" and stops the algorithm. As a result, the order 1 LCS/LRS values are "$A_3A_2$" and "$A_3A_1$", the order 2 LCS/LRS values are "$A_1A_3A_2$" and "$A_1A_3A_1$", and the order 3 LCS/LRS value is "$A_2A_1A_3A_1$", respectively.

### III. RELATED WORK

Forecasting issues can be solved using several methods, which are called soft computing. Among them, researchers often apply FTS in order to predict stock prices since stock data is a time series stream. The concept of FTS for forecasting was first proposed by Song and Chissom [6] in 1993. After that, FTS was improved by high-order relationships [7], combined with EC [4, 7] and two-factors [4, 8], which adds other indexes to help with forecasting, etc.

Moreover, Seng and Cheng [9] point out an issue that may reduce the accuracy of forecasting. Traditional FTS maps the price to the interval, which is defined by the user, and uses one/higher order logical relationships in order to represent the historical data. For example, $A_1 \rightarrow (A_1,A_3)$, and $A_2 \rightarrow (A_1,A_3)$ where the left/right hand side of the arrow are called "left/right hand side" (LHS/RHS); if the state is $A_1$, then the next state may be forecast as a number that is a weight combined with $A_1$ and $A_3$. As a result, it is neither $A_1$ nor $A_3$, and the result is not a pattern in history. Thus, they changed the mapping size of the LHS until the RHS is only matched by a subsequence. For example, the results of FTS are $A_2A_1 \rightarrow A_1$ and $A_1A_3A_2 \rightarrow A_3$. The experimental results [9] show that the accuracy of forecasting improves the forecasting of the University of Alabama's enrollments as a result of using the new mapping technique. However, the method may still be invalid for the stock market because stock price data are more complex. But it is nevertheless a good concept, which is somewhat similar to LCS/LRS. However, this being said, LCS/LRS are more powerful tools to explore past data.

Furthermore, most researchers make the decisions to use the highest and lowest prices in the historical data for the domain range of FTS, which is a little unreasonable because price mapping may cause fewer relationships of LHS/RHS. Thus, mapping should be relative rather than absolute, such as the percentage of difference on price between today with yesterday. For example, there are four situations that may cause the mapping of a FTS logical relationship to lose its accuracy, which are when the price is in a continuous rise/fall, and when it rises/falls at first and then falls/rises during the forecasting period. Take Figure 1 and Figure 2 for example, the days from 1 to 15 are the training days and the forecasting period begins after day 16. The situation presented in Figure 1 is one in which all prices rise at first then one or two rise as usual and another falls. The situation presented in Figure 2 is the reverse of Figure 1. Each day's price is represented by the blue line in Figure 1, and before day 15, is "1,2,4,3,4,6,5,7,8,10,9,12,13,13,15", and after day 16 is "16,18,17,18,19" As a result, any interval design of FTS will cause the forecasting to be limited to the lowest and highest price in history; which means that it cannot exactly forecast the price after day 16. The orange line in Figure 1 represents the price falls after day 16; the trend before day 15 may cause the FTS to forecast the future price as rising when the price is actually falling; and as a result, any interval design and logical relationship mapping of FTS cannot avoid reducing the forecasting rate. These two situations, presented in Figure 2, are the reverse of Figure 1; however, they reduce the forecasting rate in the same way as shown in Figure 1. The experimental results show us that these situations will really reduce the forecasting rate with FTS. This will be discussed further in Sec. 0.

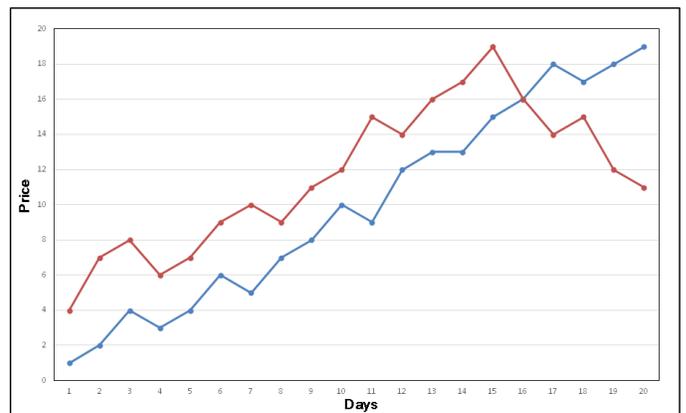

Figure 1. The price rises at first then continue rise or fall

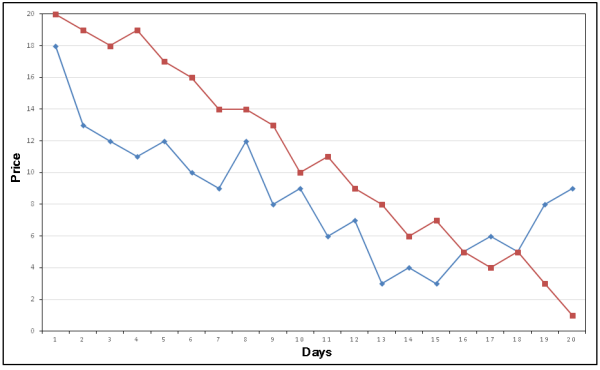

Figure 2. The price falls at first then continue fall or rise

In conclusion, this research, never proposed before, combines the features of LCS/LRS and the concept of fuzzy logic in order to explore historical data and uses a relative relationship for mapping in order to improve performance. The LCS/LRS technique is able to explore entire sets of historical data and so avoids the situations shown in Figure 1 and Figure 2. It can also provide information about real patterns occurring in history. Moreover, FTS increases the mapping rate of LCS/LRS, which returns longer LCS/LRS values through the combination of LCS/LRS and FTS. Finally, relative logical relationship mapping also increases the forecasting rate during the forecasting period.

## IV. PROPOSED METHOD

In this section, the universe design is defined first. This is relative to the range of Taiwan's stock every day and can increase the mapping rate from LCS/LRS. Subsequently, the algorithm returns the forecasting value by computing the relationship between LCS/LRS and its mapping. The algorithm includes 6 steps and is shown as follows:

*1) The universal design:* The universe of discourse refers to the extent of change of prices, where $= [D_{min}, D_{max}]$, $D_{min}$ is -7% and $D_{max}$ is 7%. These are the limits placed on the percentage of rises and falls allowed for stock fluctuation every day, which can prevent the forecasted percentage from exceeding the bound and corresponding fuzzy set. In order to make seven intervals, the set is divided into seven parts, $u_1$, $u_2, \ldots, u_6$ and $u_7$. Fourteen values are divided by seven in order to obtain the maximum and minimum values for the set. This makes $u_1= [-7, -5)$, $u_2= [-5, -3)$, $u_3= [-3, -1)$, ..., $u_6= [+3, +5)$, and $u_7= [+5, +7)$.

*2) Transform prices to pulse:* Historical data are transformed into the rises and falls of prices in percentage form, and the calculation is shown in (1).

$$\text{change extent} = \left(\frac{\text{today's price}}{\text{yesterday's price}} - 1\right) \times 100 \quad (1)$$

For example, $Price_t$ is 55.5 and $Price_{t+1}$ is 58.6, where $Price_t$ is stock price at date t. The change extant is 5.58%

*3) Fuzzify the historical data:* Fuzzify the historical data: In order to fuzzify the change extents, these are classified into corresponding sets (defined by Step 1). If the change extent is 5.58%, it belongs to $u_7$, and then is fuzzified into $A_7$ as shown in 3)The interval of each forecasting day is defined in this way.

TABLE I. THE MAPPING METHOD OF EACH DAY'S PRICE

| Date | Change extent | Universe | Interval |
|---|---|---|---|
| 20141226 | +5.58% | $u_7$ | $A_7$ |
| 20141227 | +0.65% | $u_4$ | $A_4$ |
| 20141228 | -1.31% | $u_3$ | $A_3$ |
| 20141229 | +1.20% | $u_5$ | $A_5$ |
| 20141230 | -0.55% | $u_4$ | $A_4$ |
| 20141231 | +1.02% | $u_5$ | $A_5$ |

*4) Search the LCS:* Search for patterns in the training set that match those found during the testing period. And the test period is one month before the predicted day. The match length ranges from one to the longest length that it can match. If the match length is one, then it is called degree 1. Accordingly, this search can result in many different lengths of common substrings.

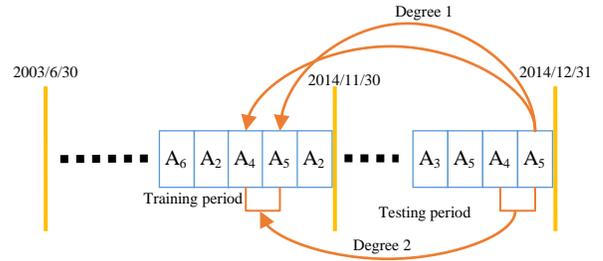

Figure 3.The method of finding same pattern

Assume that the searched for pattern is $A_3A_5A_4A_5$ during the testing period. The first set $A_5$ is searched, and the length of the same pattern is one, and so is called degree 1. One continues searching; the longest length, $A_4A_5$, also called degree 2 in Figure 3, can be obtained.

*5) Construct fuzzy logical relationships:* Based on the relationships of all degrees, a fuzzy logical relationship is constructed. Using fuzzy logical relationships enables one to obtain a fuzzy logical relationship group for all of the degrees. The percentage of rises and falls will be the midpoint of the interval. The forecasted percentage of this degree is an average of the midpoints in all intervals. The forecasted percentage of degree 1 is shown in (2), where $N_i$ is the number of the next interval if it is interval $i$, $M_i$ is the midpoint of interval $i$, and $n$ is the number of divided intervals, which has a range between two and thirty-five in our proposed model.

$$forecasted\ percentage = \frac{\sum_{i=1}^{i=n} N_i \times M_i}{\sum_{i=1}^{i=n} N_i} \quad (2)$$

TABLE II. THE TIME OF THE NEXT SET AND THE FORECASTED PERCENTAGE OF EACH DEGREE

| Pattern \ Next Interval | $A_1$ | $A_2$ | $A_3$ | $A_4$ | $A_5$ | $A_6$ | $A_7$ | Forecasted percent |
|---|---|---|---|---|---|---|---|---|
| $A_5$ | 30 | 70 | 150 | 250 | 100 | 80 | 50 | 0.082 |
| $A_4A_5$ | 25 | 50 | 100 | 200 | 80 | 60 | 40 | 0.162 |
| $A_5A_4A_5$ | 10 | 20 | 60 | 150 | 50 | 30 | 25 | 0.318 |
| $A_3A_5A_4A_5$ | 0 | 0 | 10 | 100 | 20 | 10 | 5 | 0.620 |

TABLE II. provides an example. In other words, the next interval for $A_5$ is $A_1$ and occurs 30 times, $A_2$ occurs 70 times,

etc. Moreover, the forecasted percentage is 0.082 for degree 1. The time of each degree is strictly decreasing, because longer length patterns are harder to find than shorter ones.

*6) Return to forecasted price:* Use the forecasted percentage of rises and falls for each degree to calculate the forecasted price for each degree, as shown as (3). Thus, there are different forecasted prices for all degrees. An average of all the forecasted prices will be the final forecasted price.

$$\text{forecasted price} = \text{previous price} \times \text{forecasted percent} + \text{previous price} \quad (3)$$

TABLE III. THE FORECASTED PRICE OF EACH DEGREE

| Next interval  Pattern | $A_1$ | $A_2$ | $A_3$ | $A_4$ | $A_5$ | $A_6$ | $A_7$ | Forecasted price |
|---|---|---|---|---|---|---|---|---|
| $A_5$ | 30 | 70 | 150 | 250 | 100 | 80 | 50 | 108.2 |
| $A_4A_5$ | 25 | 50 | 100 | 200 | 80 | 60 | 40 | 116.2 |
| $A_5A_4A_5$ | 10 | 20 | 60 | 150 | 50 | 30 | 25 | 131.8 |
| $A_3A_5A_4A_5$ | 0 | 0 | 10 | 100 | 20 | 10 | 5 | 162.0 |

Assume that the testing period is ……$A_3A_5A_4A_5$, that the longest length is four, and that the previous price is 100 NT. The price of all degrees is shown in TABLE III. . The final forecasted price will be the forecasted price average, 129.55.

## V. EXPERIMENT AND ANALYSIS

There remain some issues that should be solved first, which relate to the interval range and how much historical data we should use.

The parameters of the method affect the results of the experiment, and so also affect the best result. The parameters are trained in the following experiment.

The number of intervals, which can obtain the best accuracy rate in FTS, is a perplexing and confusing issue for several researchers. Therefore, the numbers of intervals are discussed in experiment 1. This experiment applies ETF from July 2014 to March 2015. The trends of ETF are able to represent Taiwan's stock market, because the constituent stock of ETF contains Taiwan's top 50 stocks.

In order to draw a comparison with Chen's method [4], the proposed forecast prices in November and December in the TAIEX from 1990 to 2004 are shown in experiments 2 and 3.

Different lengths of historical data can receive diverse results using the proposed method, and so this parameter of lengths of historical data is decided by experiment 2. The best parameter obtained from experiments 1 and 2 is used in experiment 3.

In order to draw a comparison with Chen's method [4], we use the root mean square error (RMSE) forecast accuracy rate. RMSE is defined as (4), where n denotes the number of dates that need to be forecasted.

$$\text{RMSE} = \sqrt{\frac{\sum_{i=1}^{n}(\text{forecasted price}_i - \text{actual price}_i)^2}{n}} \quad (4)$$

There is another proper tool to calculate the forecast accuracy rate, which is mean absolute percentage error (MAPE).

$$\text{MAPE} = \frac{1}{n}\sum_{i=1}^{n}\left|\frac{\text{forecasted price}_i - \text{actual price}_i}{\text{actual price}_i}\right| \times 100 \quad (5)$$

For example:

TABLE IV. THE COMPARISON OF RMSE AND MAPE

| Next day  Today price | Actual price Rise 1% | Predict price Rise 2% | RMSE | MAPE |
|---|---|---|---|---|
| 1000 | 1010 | 1020 | 10 | 0.99% |
| 10000 | 10100 | 10200 | 100 | 0.99% |

From the TABLE IV, we are able to see that the two instances differ by 1%, although the second today price exceeds the first, their error rate should be the same, like the result of MAPE. But just because the second day's price is greater than the first, it this causes a rise in RMSE. So when the RMSE is big, we cannot conclude with any certainty whether it really is a loss of prediction or whether the sample is too big.

### A. Experiment 1

How to distinguish intervals in order to achieve high prediction accuracy has been a commonly researched topic in this area since the number of interval affects the forecasted result in FTS. In this experiment, we separate U in equal length from two to thirty-one and apply an ETF from July 2014 to March 2015. Subsequently, the RMSE is calculated for each day.

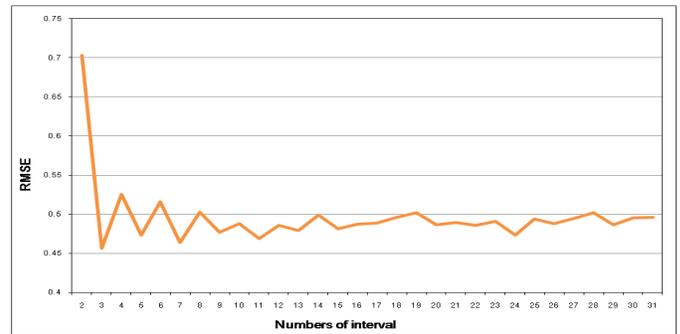

Figure 4. The RMSE of the different intervals in ETF from July 2014 to March 2015

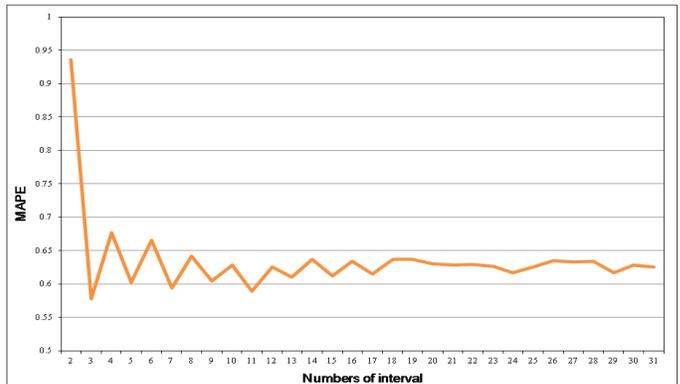

Figure 5. The MAPE of the different intervals in ETF from July 2014 to March 2015

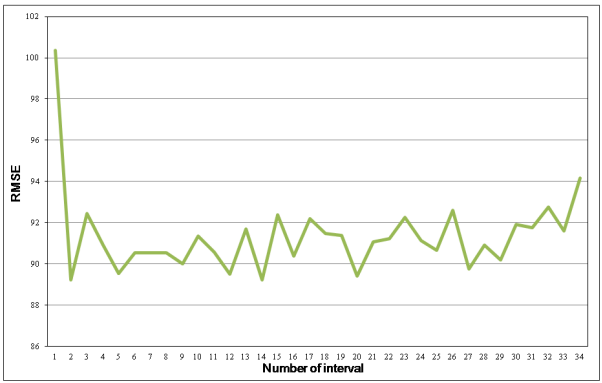
Figure 6. The RMSE of the different intervals of TAIEX from 1990 to 2004

Based on Fig. 7, 8, 9 and 10, the trend shows that odd numbers present lower RMSE values with respect to even numbers, because the bounds of odd numbers can stride across both negative and positive percentages, which makes predictions uncertain. Moreover, when the interval is 3, it can reach the smallest RMSE value in the Fig.4. Because the pulses in ETF are small, the rises and falls of stock prices move up and down mostly between $\pm 2.33\%$.

Given small fluctuations in ETF, this can also explain how the average longest common substring occurs when the interval is 3 in Fig. 11.

### B. Experiment 2

We use different lengths of training period to predict stock price in November and December 2004 in TAIEX; where this data was taken forward from 2004. Furthermore, we test 2004 because it is the last forecasted year in Chen's method [4]. When the length is 1, this means that the training period only took place in 2003.

When the length of the training period is longer, the RMSE lowers because more historical data exist for forecasting price. So this experiment shows that it is better to use historical data for as long as possible.

### C. Experiment 3

Chen's method [4] is an academic authority in the field of fuzzy logic. Chen proposed several meaningful fuzzy methods for forecasting price in TAIEX, and these reformed methods have been shown to obtain better accuracy rates than previous methods. There is no denying that Chen is the most important scholar in the field of fuzzy forecasting, and so the method proposed here is compared with his method as proposed in 2013.

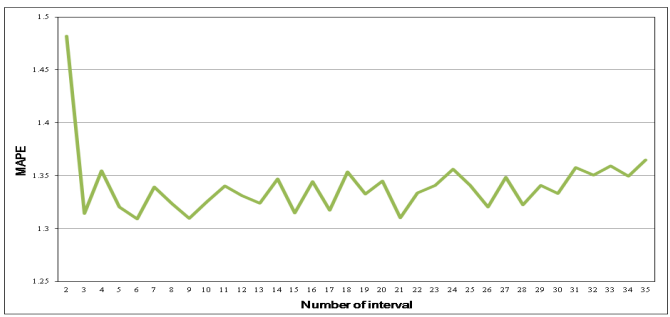
Figure 12. The MAPE of the different intervals of TAIEX from 1990 to 2004

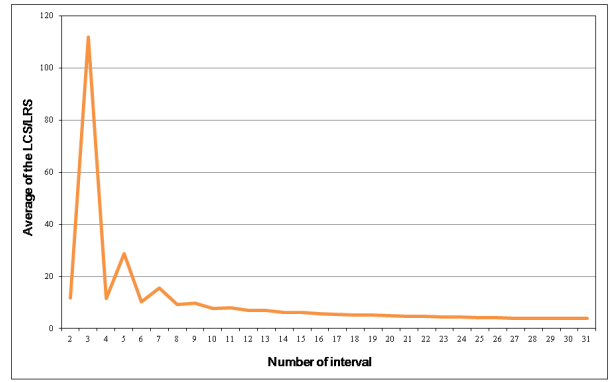
Figure 13. The average length of the different intervals from July 2014 to March 2015

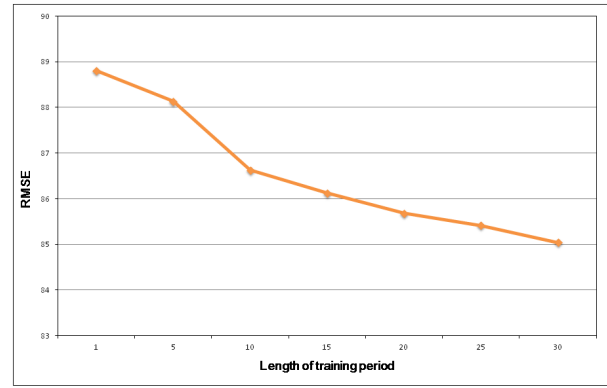
Figure 14. The RMSE of 2004 with different lengths of training period

The results are shown in TABLE V, VI. Although the RMSE values are much smaller than those obtained using Chen's method in 2000 and 2004, the calculations made by the proposed method are much easier to run than theirs, and the results obtained in other years are nearly the same as those obtained from their method.

TABLE V, VI shows the comparison with Chen's model from 1990 to 2004.

TABLE V. A COMPARISON OF THE RMSE VALUES FOR THE DIFFERENT METHODS

| | Methods | 1999 | 2000 | 2001 | 2002 | 2003 | 2004 | AVG. |
|---|---|---|---|---|---|---|---|---|
| [4] | Factor1 | 102.34 | 131.25 | 113.62 | 65.77 | 52.23 | 56.16 | 86.89 |
| | Factor2 | 102.11 | 131.3 | 113.83 | 66.45 | 52.83 | 54.17 | 86.78 |
| | Factor3 | 103.52 | 131.36 | 112.55 | 66.23 | 53.2 | 55.36 | 87.03 |
| | Ours (interval=3) | 102.8 | 117.58 | 114.64 | 65.92 | 52.55 | 51.28 | 84.12 |

Note: factor 1 is Dow Jones, factor 2 is NASDAQ, and factor 3 is M1b.

TABLE VI. A COMPARISON OF THE RMSE VALUES FOR THE DIFFERENT METHODS

| | Methods | 1990-1993 | 1993-1996 | 1996-1999 | 1999-2002 | 2001-2004 |
|---|---|---|---|---|---|---|
| [4] | Factor1 | 92.32 | 75.26 | 101.7 | 103.25 | 71.95 |
| | Factor2 | 92.33 | 74.75 | 101.375 | 103.42 | 71.82 |
| | Factor3 | | | | 103.42 | 71.84 |
| | Ours (interval=3) | 91.47 | 73.12 | 105.6 | 100.23 | 71.1 |

Note: factor1 is Dow Jones, factor2 is NASDAQ, and factor3 is M1b.

Chen's method has poor performance in 1990 and 2000. Fig. 10 and Fig. 11 are the TAIEX price at 1990 and 2000. It is obviously that it meets the condition we mention in Sec. 0. The prices in 1990 and 2000 are falling continuously from January to October, one is continuous falling at November and December, and the other is continuous rising at November and December. The two situation cause its RMSE are bigger than other years.

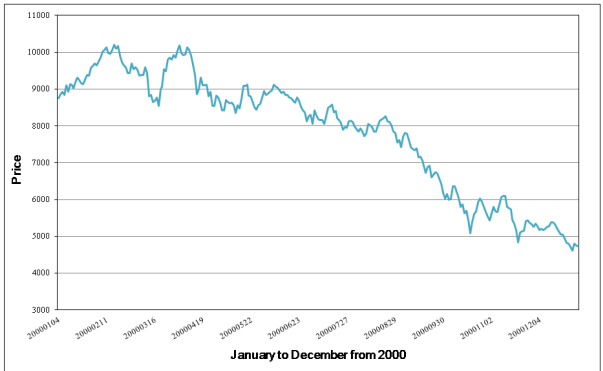

Figure 10. The price of TAIEX in 2000

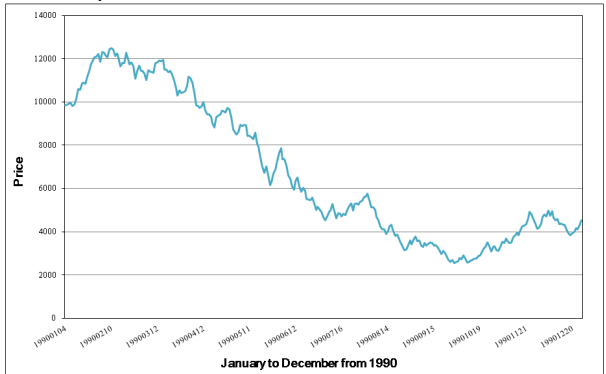

Figure 11. The price of TAIEX in 2000

VI. CONCLUSION AND FUTURE WORK

Our research uses LCS/LRS to identify historical stock price trends. However, because prices cannot be exactly matched, we fuzzify these prices into a fuzzy set. The traditional application on FTS uses the highest and lowest prices in the historical data as the upper and lower bounds in the fuzzy set. This can cause the predicted price to overflow the boundary and lose forecast accuracy. The proposed method therefore uses the percentage of rises and falls of the stock price as fuzzy intervals. Taiwan's stock markets, such as ETF and TAIEX, have a limit of 7% imposed on rising/falling, and so our proposal need not take this problem into consideration. The experimental results show that when the interval is three and of equal length, performance is better than other numbers of intervals and that odd numbers present lower RMSE values with respect to even numbers. Our proposed method is easy to implement and does not provide bad results. In the future certain parameters could be adjusted. First, perhaps we can use the percentage of rises and falls every two or three days…etc. Second, in our proposed method every degree is assigned an average weight, but perhaps we can adjust this weight; namely, perhaps the results will be better if higher degrees are assigned more weight. Finally, when referring to historical data, recent data may be adopted more easily than relatively longer data.